\documentstyle[12pt]{article}
\begin{document}
\def\bea{\begin{eqnarray}}
\def\eea{\end{eqnarray}}
\title{\bf {Holographic Thermodynamic on the Brane in Topological Reissner-Nordstr\"om
 de Sitter Space }}
 \author{$^{1,2,3}$
M.R. Setare  \footnote{E-mail: rezakord@yahoo.com},$^{1,3}$R.
Mansouri \footnote{ E-mail:mansouri@sharif.edu }
\\
 {$^{1}$ Department of Physics, Sharif
University of Technology, Tehran, Iran
 }\\
{$^{2}$ Department of Science, Physics group, Kurdistan
University, Sanandeg, Iran }\\{$^{3}$Institute for Theoretical
Physics and Mathematics, Tehran, Iran }}
\maketitle
\begin{abstract}

We consider the brane universe in the bulk background of the
 topological Reissner-Nordstr\"om de Sitter black holes.
 We show that the thermodynamic quantities (including entropy)
 of the dual CFT take usual special forms expressed in terms
 of Hubble parameter and its time derivative at the moment ,
  when the brane crosses the black hole horizon or the cosmological horizon.
 We obtain the generalized Cardy-Verlinde formula for the
CFT with an charge and cosmological constant, for any values of
the curvature parameter $k$ in the Friedmann equations.

 \end{abstract}
\newpage

 \section{Introduction}
The holographic duality which connects $n+1$-dimensional gravity
in Anti-de Sitter (AdS) background with $n$-dimensional conformal
field theory (CFT) has been discussed vigorously for some
years\cite{AdS}. But it seems that we live in a universe with a
positive cosmological constant which will look like de Sitter
space--time in the far future. Therefore, we should try to
understand quantum gravity or string theory in de Sitter space
preferably in a holographic way.. Of course, physics in de Sitter
space is interesting even without its connection to the real
world; de Sitter entropy and temperature have always been
mysterious aspects of quantum gravity\cite{GH}.\\
While string theory successfully has addressed the problem of
entropy for black holes, dS entropy remains a mystery. One reason
is that the finite entropy seems to suggest that the Hilbert space
of quantum gravity for asymptotically de Sitter space is finite
dimensional, \cite{{Banks:2000fe},{Witten:2001kn}}.
 Another, related, reason is that the horizon and entropy in
de Sitter space have an obvious observer dependence. For a black
hole in flat space (or even in AdS) we can take the point of view
of an outside observer who can assign a unique entropy to the
black hole. The problem of \ what an observer venturing inside the
black hole experiences, is much more tricky and has not been given
a satisfactory answer within string theory. While the idea of
black hole complementarity provides useful clues, \cite
{Susskind}, rigorous calculations are still limited to the
perspective of the outside observer. In de Sitter space there is
no way to escape the problem of the observer dependent entropy.
This contributes to the difficulty of de Sitter space.\\
 Recently
much attention has been paid for the duality between de Sitter
(dS) gravity and CFT by the analogy of the AdS/CFT
correspondence\cite{AS,dS,CI,set} (for a very good review see also
\cite{odi}) , because the isometry of $n+1$-dimensional de Sitter
space, $SO(n+1,1)$, exactly agrees with the conformal symmetry of
$n$-dimensional Euclidean space. Thus it might be natural to
expect the correspondence between $n+1$-dimensional gravity in de
Sitter space and $n$-dimensional Euclidean CFT (the dS/CFT
correspondence). Moreover the holographic principle between the
radiation dominated Friedmann-Robertson-Walker (FRW) universe in
$n$-dimensions and same dimensional CFT with a dual
$n+1$-dimensional AdS description was studied in \cite{EV}.
Especially, we can see the correspondence between black hole
entropy and the entropy of the CFT which is derived by making the
appropriate identifications for
FRW equation with the  Cardy-Verlinde formula.\\
 In this paper we consider the brane universe in the bulk background of the
 topological Reissner-Nordstr\"om de Sitter (TRNdS) black holes.
 We show that the thermodynamic quantities (including entropy)
 of the dual CFT take usual special forms expressed in terms
 of Hubble parameter and its time derivative at the moment ,
  when the brane crosses the black hole horizon or the cosmological horizon.
 We obtain the generalized Cardy-Verlinde formula for the
CFT with an charge and cosmological constant , for any values of
the curvature parameter $k$ in the Friedmann equations.

\section{FRW equations in the background of TRNdS Black Holes}
The topological Reissner-Nordstr\"om dS black hole solution in
$(n+2)$-dimensions has the following form
\begin{eqnarray}
&& ds^2 = -f(r) dt^2 +f(r)^{-1}dr^2 +r^2 \gamma_{ij}dx^{i}dx^{j}, \nonumber \\
&&~~~~~~ f(r)=k -\frac{\omega_n M}{r^{n-1}} +\frac{n \omega_n^2
Q^2}{8(n-1) r^{2n-2}}
     -\frac{r^2}{l^2},
\end{eqnarray}
where
\begin{equation}
\omega_n=\frac{16\pi G}{n\mbox {Vol}(\Sigma)},\hspace{1cm}\phi
=-\frac{n}{4(n-1)}\frac{\omega_n Q}{r^{n-1}},
\end{equation}
where $Q$ is the electric/magnetic charge of Maxwell field, $M$ is
assumed to be a positive constant , $l$ is the curvature radius of
de Sitter space, $\gamma_{ij}dx^idx^j$ denotes the line element of
an $n$-dimensional hypersurface $\Sigma_k$ with the constant
curvature $n(n-1)k$ and its volume $V(\Sigma_k)$. $\Sigma_k$ is
given by spherical ($k=1$), flat ($k=0$), hyperbolic  $(k=-1)$,
$\phi$ is the elctrostatic potential related to the charge $Q$.
When $k=1$, the metric Eq.(1) is just the Reissner-Nordstr\"om-de
Sitter solution. For general $M$ and $Q$, the equation $f(r)=0$
may have four real roots. Three of them are real, the largest on
is the cosmological horizon $r_{c}$, the smallest is the inner
(Cauchy) horizon of black hole, the middle one is the outer
horizon $r_{+}$ of the black hole. And the fourth is negative and
has no physical meaning. The case $M=Q=0$ reduces
to the de Sitter space with a cosmological horizon $r_{c}=l$.\\
When $k=0$ or $k=-1$, there is only one positive real root of
$f(r)$, and this locates the position of cosmological horizon
$r_{c}$.\\
In the case of $k=0$, $\gamma_{ij}dx^{i}dx^{j}$ is an
$n-$dimensional Ricci flat hypersurface, when $M=Q=0$ the solution
Eq.(1) goes to pure de Sitter space
\begin{equation}
ds^{2}=\frac{r^{2}}{l^{2}}dt^{2}-\frac{l^{2}}{r^{2}}dr^{2}+r^{2}dx_{n}^{2}
\end{equation}
in which $r$ becomes a timelike coordinate.\\
When $Q=0$, and $M\rightarrow -M$ the metric Eq.(1)is the TdS
(Topological de Sitter) solution \cite{{cai93},{med}}, which have
a cosmological horizon and a naked singularity.\\
 For the purpose of getting the Friedmann-Robertson-Walker(FRW) metric, we impose
the following condition\cite{EV},
\begin{equation}
{1\over{f(r)}}\left({{dr}\over{d\tau}}\right)^2-f(r)\left({{dt}\over
{d\tau}}\right)^2=-1,
\end{equation}
which leads to a timelike brane.
 Substituting Eq.(4) into
the  TRNdS solution  Eq.(1), one has the induced brane metric
which takes FRW form
\begin{equation}
ds^2=-d\tau^2+r^2(\tau)\gamma_{ij}dx^idx^j,
\end{equation}
the equation of motion of the brane is given by\cite{SV}
 \begin{equation}
 {\cal K}_{ij}=\frac{\sigma}{n}h_{ij},
  \end{equation}
where ${\cal K}_{ij}$ is the extrinsic curvature, and $h_{ij}$ is
the induced metric on the brane, $\sigma$ is  the brane tension.
The extrinsic curvature, ${\cal K}_{ij}$, of the brane can be
calculated and expressed in term of function $r(\tau)$ and
$t(\tau)$. Thus one rewrites the equations of motion (6) as
\begin{equation}
{{dt}\over{d\tau}}=\frac{\sigma r}{f(r)} ,
\end{equation}
 Using Eqs.(4,7), we can drive FRW equation with $H=\frac{\dot{r}}{r}$,
\begin{equation}
H^2=-\frac{f(r)}{r^2}+ \sigma^2={{\omega_{n}M}\over
r^{n+1}}-{{nw^2_{n}Q^2}\over{8(n-1)r^{2n}}} -{k\over
r^2}+\frac{1}{l^{2}}+\sigma^2,
\end{equation}
where, $H$ is the Hubble parameter. Here we cannot make any
fine-tuning to obtain a flat brane. We have thus  an effectively
de Sitter brane. Making use of the fact that the metric for the
boundary CFT can be determined only up to a conformal factor, we
rescale the boundary metric for the CFT to be of the following
form
\begin{equation}
ds^2_{CFT}=\lim_{r\to\infty}\left[{l^2\over r^2}ds^2_{n+2}\right]
=-dt^2+l^2\gamma_{ij}dx^idx^j.
\end{equation}
Then the thermodynamic relations between the boundary CFT and the
bulk TRNdS are given by
\begin{eqnarray}
E_{CFT}&=&M{l\over r},\ \ \ \ \ \ \ \ \ \ \Phi_{CFT}=\phi{l\over
r},
 \cr T_{CFT}&=&{T_{TRNdS}}{l\over r},\ \ \ \ \ \ \ S_{CFT}=S_{TRNdS}
\end{eqnarray}
where black hole horizon Hawking temperature $T_{TRNdS}^{b}$ and
entropy $S_{TRNdS}^{b}$ are given by
\begin{eqnarray}
 && T_{TRNdS}^{b}=\frac{f'(r_{+})}{4\pi} =\frac{1}{4\pi r_+} \left((n-1) -(n+1)\frac{r_+^2}{l^2}
   -\frac{n\omega_n^2 Q^2}{8 r_+^{2n-2}}\right), \nonumber \\
&& S_{TRNdS}^{b} =\frac{r_+^n\mbox{Vol}(\Sigma)}{4G},
\end{eqnarray}
where $r=r_{+}$ is black hole horizon and
$v_{+}=r_{+}^{n}Vol(\Sigma)$ is area of it in $(n+2)-$dimensional
asymptotically dS space. The Hawking temperature $T_{TRNdS}^{c}$
and entropy $S_{TRNdS}^{c}$ associated with the cosmological
horizon are
\begin{eqnarray}
 && T_{TRNdS}^{c}=\frac{-f'(r_{c})}{4\pi} =\frac{1}{4\pi r_c} \left(-(n-1)k +(n+1)\frac{r_c^2}{l^2}
    +\frac{n\omega_n^2 Q^2}{8 r_c^{2n-2}}\right), \nonumber \\
&& S _{TRNdS}^{c}=\frac{r_c^n\mbox{Vol}(\Sigma)}{4G},
\end{eqnarray}
where $V_{c}=r_{c}^{n}Vol(\Sigma)$ is area of the cosmological
horizon.In terms of the energy density $\rho_{CFT}=E_{CFT}/V$, the
pressure $p_{CFT}=\rho_{CFT}/n$, the charge density
$\rho_{QCFT}=Q/V$ and the electrostatic potential
$\Phi_{CFT}=\phi{l\over r}$ of the CFT within the volume
$V=r^n{\rm Vol}(\Sigma)$, the first Friedmann equation take the
following form
\begin{equation}
H^2={{16\pi G}\over{n(n-1)}}\left(\rho_{CFT}-{1\over
2}\Phi\rho_{QCFT}\right) -{k\over
r^2}+\frac{2\Lambda_{n+1}}{n(n+1)},
\end{equation}
with the positive cosmological constant
$\Lambda_{n+1}=\frac{n(n+1)}{2}(1/\ell^2 + \sigma^2)$, and
$G=\frac{(n-1)G_{n+2}}{l}$. Taking the $\tau-$derivative of
Eq.(8), we obtain the second Friedmann equation
\begin{equation}
\dot{H}=-{{n+1}\over
2}{{\omega_{n}M}\over{r^{n+1}}}+{{n^2w^2_{n}Q^2}
\over{8(n-1)r^{2n}}}+{k\over r^2}.
\end{equation}
Similar to the first Friedmann equation we can rewrite the second
equation as following
\begin{equation}
\dot{H}=-{{8\pi
G}\over{n-1}}\left(\rho_{CFT}+p_{CFT}-\Phi_{CFT}\rho_{QCFT}\right)+{k\over
r^2}, \label{frd2}
\end{equation}
 So, the
cosmological evolution is determined by the energy density
$\rho_{CFT}$ and the pressure $p_{CFT}$, electric potential energy
and cosmological constant. The Friedmann equations (13,15) can be
respectively put into the following forms resembling thermodynamic
formulas of the CFT,
\begin{equation}
S_H={{2\pi}\over
n}r\sqrt{E_{BH}[2(E_{CFT}+E_{\Lambda}-\textstyle{1\over
2}\Phi_{CFT} Q)-kE_{BH}]},
\end{equation}
\begin{equation}
kE_{BH}=n(E_{CFT}+pV-\Phi_{CFT} Q-T_HS_H),
\end{equation}
in terms of the Hubble entropy $S_H$ and the Bekenstein-Hawking
energy $E_{BH}$  the cosmological energy $E_{\Lambda}$ and the
Hubble temperature $T_H$, where
\begin{equation}
S_H\equiv (n-1){{HV}\over{4G}}, \ \ \ \ \
E_{\Lambda}\equiv\frac{\Lambda (n-1)V}{8\pi G (n+1)}\ \ \ \ \ \
E_{BH}\equiv n(n-1){V\over{8\pi G r^2}}, \ \ \ \ \  T_H\equiv
-{\dot{H}\over{2\pi H}}.
\end{equation}
 Now we study thermodynamics of the CFT at the moment when the brane
crosses the black hole horizon $r=r_+$ in the case $k=1$, and
crosses the cosmological horizon $r=r_{c}$ for the cases $k=0$ and
$k=-1$ defined as the largest root of $f(r)=0$,
\begin{eqnarray}
k -\frac{\omega_n M}{r^{n-1}} +\frac{n \omega_n^2 Q^2}{8(n-1)
r^{2n-2}}  -\frac{r^2}{l^2}=0.
\end{eqnarray}
From Eqs.(8,19), after setting $\sigma=\frac{1}{l}$, we have
\begin{equation}
H^2={1\over l^2}\ \ \ \ \ \ \ {\rm at}\ \ \ \ \ \ r=r_{+,c}.
\end{equation}
The total entropy $S_{CFT}$ of the CFT remains constant, but the
entropy density,
\begin{equation}
s\equiv{S_{CFT}\over
V}=\frac{r_{+,c}^{n}Vol(\Sigma)}{4G_{n+2}r^{n}Vol(\Sigma)}=(n-1){r^n_{+,c}\over{4Glr^n}},
\end{equation}
varies with time.  Using Eq. (20), we see that $s$ at $r=r_{+,c}$
can be expressed in terms of $H$ as following
\begin{equation}
s=(n-1){H\over{4G}}\ \ \ \ \ \ \ {\rm at}\ \ \ \ \ \ r=r_{+,c},
\end{equation}
then we have
\begin{equation}
S=S_H\ \ \ \ \ \ \ {\rm at}\ \ \ \ \ \ r=r_{+,c}
\end{equation}
Using  the formula $H^2=\sigma^2-f(r)/r^2$ that follows from Eq.
(8), we see that the CFT temperature $T_{CFT}=
f^{\prime}(r_{+})l/(4\pi r_{+})$, at the moment when the brane
crosses the black hole horizon,  can be expressed in terms of $H$
and $\dot{H}$ in the following way
\begin{equation}
T_{CFT}=-{\dot{H}\over{2\pi H}}\ \ \ \ \ \ \ {\rm at}\ \ \ \ \ \
r=r_{+}.
\end{equation}
In the cases $k=0$ and $k=-1$, at the moment when the brane
crosses the cosmological horizon ,the CFT temperature $T_{CFT}=
-f^{\prime}(r_{c})l/(4\pi r_{c})$ can be expressed as the
following
\begin{equation}
T_{CFT}= {\dot{H}\over{2\pi H}}\ \ \ \ \ \ \ {\rm at}\ \ \ \ \ \
r=r_{c}.
\end{equation}
Using Eqs.(17,23,24)in the case $k=1$ we can write
\begin{equation}
E_C=E_{BH}\ \ \ \ \ \ \ {\rm at}\ \ \ \ \ \ r=r_{+}
\end{equation}
where $E_C$ is the Casimir energy defined as
\begin{equation}
E_C\equiv n(E_{CFT}+p_{CFT}V-\Phi_{CFT}Q-TS).
\end{equation}
If we redefine the Hubble temperature as $T_H\equiv
{\dot{H}\over{2\pi H}}$ for the cases $k=0$ and $k=-1$, then we
can write
\begin{equation}
E_C=kE_{BH}\ \ \ \ \ \ \ {\rm at}\ \ \ \ \ \ r=r_{+,c}
\end{equation}
 Therefore, for the case of topological Reissner-Nordstr\"om de
Sitter black holes, thermodynamic quantities of the CFT can be
expressed in terms of the Hubble parameter and its time
derivative, when the brane crosses the black hole horizon in the case $k=1$ and crosses the
cosmological horizon in the cases $k=0$, $k=-1$ .\\
We can rewrite Eq.(16) in term of the Casimir energy $E_C$, by
using Eqs.(23,28)
\begin{equation}
S=\frac{2\pi
r}{n}\sqrt{|\frac{E_{C}}{k}|(2(E_{CFT}+E_{\Lambda}-1/2\phi_{CFT}Q)-E_{C})},
\end{equation}
this is the generalized Cardy-Verlinde formula for the CFT with
cosmological constant $\Lambda$ and charge $Q$. Therefore the
generalized Cardy-Verlinde Eq.(29) for boundary CFT, coincides
with the cosmological Cardy formula Eq.(16) when the brane crosses
the black hole horizon $r=r_{+}$ or cosmological horizon
$r=r_{c}$.

  \section{Conclusion}
One of the striking results for the dynamic dS/CFT correspondence
is that the Cardy-Verlinde's formula on the CFT-side coincides
with the Friedmann equation in cosmology when the brane crosses
the horizon $r=r_{+}$ of the topological Reissner-Nordstr\"om
black hole. This means that the Friedmann equation knows the
thermodynamics of the CFT. In this paper we have considered the
brane universe in the bulk background of the
 topological Reissner-Nordstr\"om de Sitter black holes.
We have shown that in TRNdS space in contrast with TRNAdS space we
cannot make any fine-tuning to obtain a flat brane. We have thus
an effectively de Sitter brane. We have shown that thermodynamic
quantities of the CFT can be expressed in terms of the Hubble
parameter and its time derivative, when the brane crosses the
black hole horizon or cosmological horizon respectively for the
case $k=1$ and the cases $k=0$, $k=-1$. We obtain the generalized
Cardy-Verlinde formula for the CFT with an charge and cosmological
constant , for any values of the curvature parameter $k$ in the
Friedmann equations.

\end{document}